\documentclass{an}
\usepackage{graphicx}
\usepackage{times}
\Volume{01}              
\Year{2003}              
\Month{1}                
\Pagespan{1}{4}      

\begin{document}
\lhead[\thepage]{X. Barcons et al: The XMM-Newton SSC MS}
\rhead[Astron. Nachr./AN~{\bf XXX} (200X) X]{\thepage}
\headnote{Astron. Nachr./AN {\bf 32X} (200X) X, XXX--XXX}

\title{The XMM--Newton Survey Science Centre Medium Sensitivity Survey\thanks{Based partly
on observations obtained with XMM-Newton, an ESA science mission with
instruments and contributions directly funded by ESA member states and
the USA (NASA)}}

\author{X. Barcons\inst{1}\and F.J. Carrera\inst{1} \and
M.T. Ceballos\inst{1} \and S. Mateos\inst{1} \and M.J. Page\inst{2} \and I. P\'erez-Fournon\inst{3} \and M.G. Watson\inst{4} \and the XMM-Newton Survey
Science Centre}
\institute{
Instituto de F\'\i sica de Cantabria (CSIC-UC), 39005 Santander, Spain \and
Mullard Space Science Laboratory, UCL, Holmbury St Mary, Dorking, Surrey RH5 6NT, UK \and
Instituto de Astrof\'\i sica de Canarias, 38200 La Laguna, Tenerife, Spain \and
Department of Physics and Astronomy, University of Leicester, LE1 7RH, UK}

\date{Received {\it date will be inserted by the editor}; 
accepted {\it date will be inserted by the editor}} 

\abstract{We present preliminary results on the XMM-Newton Survey
Science Centre medium sensitivity survey, with 0.5-4.5 keV flux limit
$2\times 10^{-14}\, {\rm erg}\, {\rm cm}^{-2}\, {\rm s}^{-1}$. At
present, 19 fields have been examined with a total of 239 X-ray
sources. Identifications for about 2/3 of these reveal that type 1 AGN
dominate, but type 2 AGN, absorption-line galaxies and clusters as well as
stars are also present. We also discuss on a few X-ray selected
Broad-Absorption-Line QSOs found in this survey
}

\correspondence{barcons@ifca.unican.es}

\maketitle

\section{Introduction}

After more than 2 years of scientific operations, XMM-Newton is
building up smoothly an unsurpassed serendipitous sky survey of X-ray
sources at an approximate rate of 50,000 new sources per year.  The
XMM-Newton Survey Science Centre is conducting various activities to
characterise and, eventually, fully exploit the content of this survey
(Watson et al 2001).  This will be achieved via an imaging programme,
whereby many XMM-Newton target fields are being imaged in the optical
(often in multiple bands) and near-infrared, and a core programme where
large samples of carefully selected X-ray sources are fully identified
via archival searches and optical spectroscopy.

This paper presents the current status of the XMM SSC Medium Survey
(XMS) at high galactic latitudes.  The XMS has the goal of
characterising the 0.5-4.5 keV X-ray selected source population down
to a source density of $\sim 100\, \deg^{-2}$, which corresponds to a
flux limit of $2\times 10^{-14}\, {\rm erg}\, {\rm cm^{-2}}\, {\rm
s^{-1}}$. A first compilation of 29 X-ray sources in two XMM-Newton
fields has been presented in Barcons et al (2002). The XMS will bridge
the gap between the deep surveys (e.g., Hasinger et al 2001), aimed at
exploiting the most distant Universe, and the
shallower bright sample (Della Ceca et al 2002) which is meant to
characterise the local and high luminosity X-ray source populations.  

\section{The XMM-Newton SSC Medium Survey (XMS)}

The XMS is being built by using EPIC pn images of XMM-Newton
observations fulfilling a number of criteria: galactic latitude $\mid
b\mid > 20\, \deg$, good-time-intervals exceeding $\sim 10$~ks
(typically $\sim 20$ ks), no bright ($\sim 10^{mag}$) stars in the
field and moderate Galactic column density ($<10^{21}\, {\rm
cm}^{-2}$).  We used XMM-Newton data where the corresponding PI had
granted access to the Survey Science Centre to exploit the
serendipitous content of the fields. The targets of the observations
have been obviously excluded, together with a region around them (that
depending on the brightness and/or extension of the X-ray target).

The data had been processed by the Survey Science Centre
pipeline with SAS version 5.2.  We plan to update the current source
lists with the more reliable SAS version 5.3.3 processed data
products, but this will happen not earlier than the beginning of
2003. That new version of the source lists should have better
calibrated source fluxes and improved positions.

At present we have only included fields visible from the Observatorio
del Roque de Los Muchachos in the island of La Palma (Spain), where
all the spectroscopic identifications have been obtained. There is an
on-going complementary effort in the South that will ultimately double
the size of this sample.

In the present paper we discuss a sample of X-ray sources found in 19
XMM-Newton EPIC pn fields. All XMM-Newton observations, except for
those referring to one particular field, are now public.  The EPIC pn
source lists were searched for sources having a flux in the 0.5-4.5
keV band, as obtained from the sum of the 0.5-2 and 2-4.5 keV standard
subbands, exceeding $2\times 10^{-14}\, {\rm erg}\, {\rm cm}^{-2}\,
{\rm s}^{-1}$. Given the rather long exposure times, the likelihood
associated to these sources is in most cases significantly larger than
the overall limit imposed to the standard 5-band search of 16 (see,
e.g., Barcons et al 2002 for details). The sample contains a total of
239 sources, covering approximately $\sim 2.3\, \deg^2$ of sky, with a
source density of $\sim 100\, {\rm sources}\, \deg^{-2}$.

\section{Optical observations}

All but one of the 19 XMM-Newton fields were imaged with the Wide
Field Camera (WFC) on the INT on the island of La Palma in various
filters. The remaining field is part of the Subaru deep survey and was
imaged by the Subaru telescope.  For the 18 fields observed with the
INT, 16 were imaged in the 4 filters g'(SDSS), r'(SDSS), i'(SDSS) and
Z (Gunn), one with r' and i' and one with i' only.  The depth of the
optical images varied slightly as sky conditions were different, but
on average the i'-band image had a limiting magnitude $\sim
23^{mag}$. Images were astrometrically calibrated against the USNO A.2
catalogue, with residuals $< 0.3\, {\rm arcsec}$, and photometrically
calibrated to zeroth order using nightly average extinction constants and
observations of standard stars. Calibrated optical images for a large
number of XMM-Newton fields can be downloaded from the XMM-Newton SSC
site (http://xmmssc-www.star.le.ac.uk) and will ultimately enter the
XMM-Newton Science Archive maintained by ESA.

We used either the i'-band or the r'-band images to search for
candidate counterparts around each X-ray source.  The algorithm
searched within 5 standard deviations in the {\it statistical}
positioning error of XMM-Newton (typically 1 arcsec error) or 5 arcsec,
whichever is larger.  This last criterion was included in order to
account for any residual systematics in the XMM-Newton attitude orbit
control system astrometric solution
of the X-ray data. With the newer SAS version 5.3.3 processed
data products, the astrometry of the source lists is automatically
corrected via a cross-correlation with a number of catalogues,
resulting in very small residual systematics ($<2\, {\rm arcsec}$).

The majority of the XMS sources had a single optical candidate
counterpart in the optical images.  A few of them had more than one
and about 11\% showed either faint (i.e. close to our detection limit
of $i'\sim 23$) or no candidate counterpart within our sensitivity
limits. It is still possible that a few of these sources are not real
X-ray sources, but this will have to await to the new SAS 5.3.3 source
lists to be confirmed.

Optical candidates were spectroscopically observed within the
AXIS\footnote{For more information on AXIS (An XMM-Newton
International Survey), visit our page http://www.ifca.unican.es/\~{}xray/AXIS}
programme at La Palma. AXIS is an international time programme which
was granted a total of 85 telescope nights spread over April 2000 -
April 2002 distributed in the 4 major telescopes at the Observatorio
del Roque de Los Muchachos: the Isaac Newton Telescope (2.5m, used for
imaging), the Nordic Optical Telescope (2.6m, used for long-slit
spectroscopy), the Telescopio Nazionale Galileo (3.5m, used mostly for
long-slit spectroscopy) and the William Harschel Telescope (4.2m, used
for both fibre and long-slit spectroscopy). 

\section{Identifications}

Out of the 239 X-ray sources in the sample, we have successfully
identified 134 (56\%) and a further 25 (10\%) are identified with some
residual uncertainty.  These uncertainties typically refer to the
redshift of the source which in some cases, particularly in those
observed with the fibre spectrograph which had shorter wavelength
coverage ($\sim 3000$\AA), was drawn from one emission line only. As
previously said a further 27 sources (11\%) had either very faint
(barely visible) or no optical counterpart down to magnitude ${\rm
i}'\sim 23$.  There is a final lot of 53 sources (22\%) which have not
been spectroscopically observed or whose optical spectra are of poor
quality.  A significant fraction of these (about 1/3) are too faint
for spectroscopic identification in 4m-class telescopes and will
therefore require 8-10m class telescope observations.

\begin{table}[h]
\caption{Breakdown of identified sources in the XMS.}
\label{ids}
\begin{tabular}{l r r}\hline
Source class & Number & \% of ids.\\ 
\hline
BLAGN& 110 & 69.2 \\
NELG &  21 & 13.2 \\
Stars & 18 & 11.3\\
Galaxies \& Clusters & 9 & 5.7\\
BL Lacs & 1 & 0.6\\
\hline
\end{tabular}
\end{table}

The breakdown of identified sources is presented in table
\ref{ids}. At a variance with other surveys, here we use a purely
optical classification scheme: Broad-line Active Galactic Nuclei
(BLAGN), Narrow Emission Line Galaxies (NELG), Galaxies and Clusters
for those not showing any emission lines (as some of the
absorption-line galaxies may lie in clusters unltimately responsible
for most of the X-ray emission), Stars and BL Lacs.

\begin{figure}[h]
\resizebox{\hsize}{!}
{\includegraphics{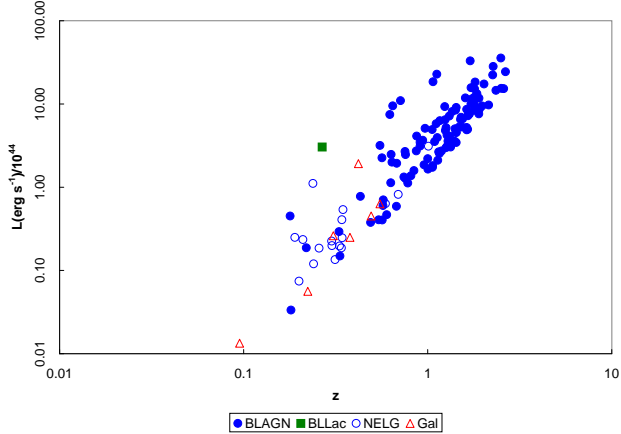}}
\caption{Luminosity-redshift diagram for the extragalactic sources
identified.}
\label{Lz}
\end{figure}

Figure \ref{Lz} presents the luminosity redshift diagram for the
extragalactic identified sources.   As it can be seen,
all NELGs in this diagram have an X-ray luminosity in excess of
$10^{42}\, {\rm erg}\, {\rm s}^{-1}$ and therefore there is no doubt
they harbour an active nucleus. In a more physical scheme,
they would be classified as narrow-line AGNs.

\begin{figure}[h]
\resizebox{\hsize}{!}
{\includegraphics{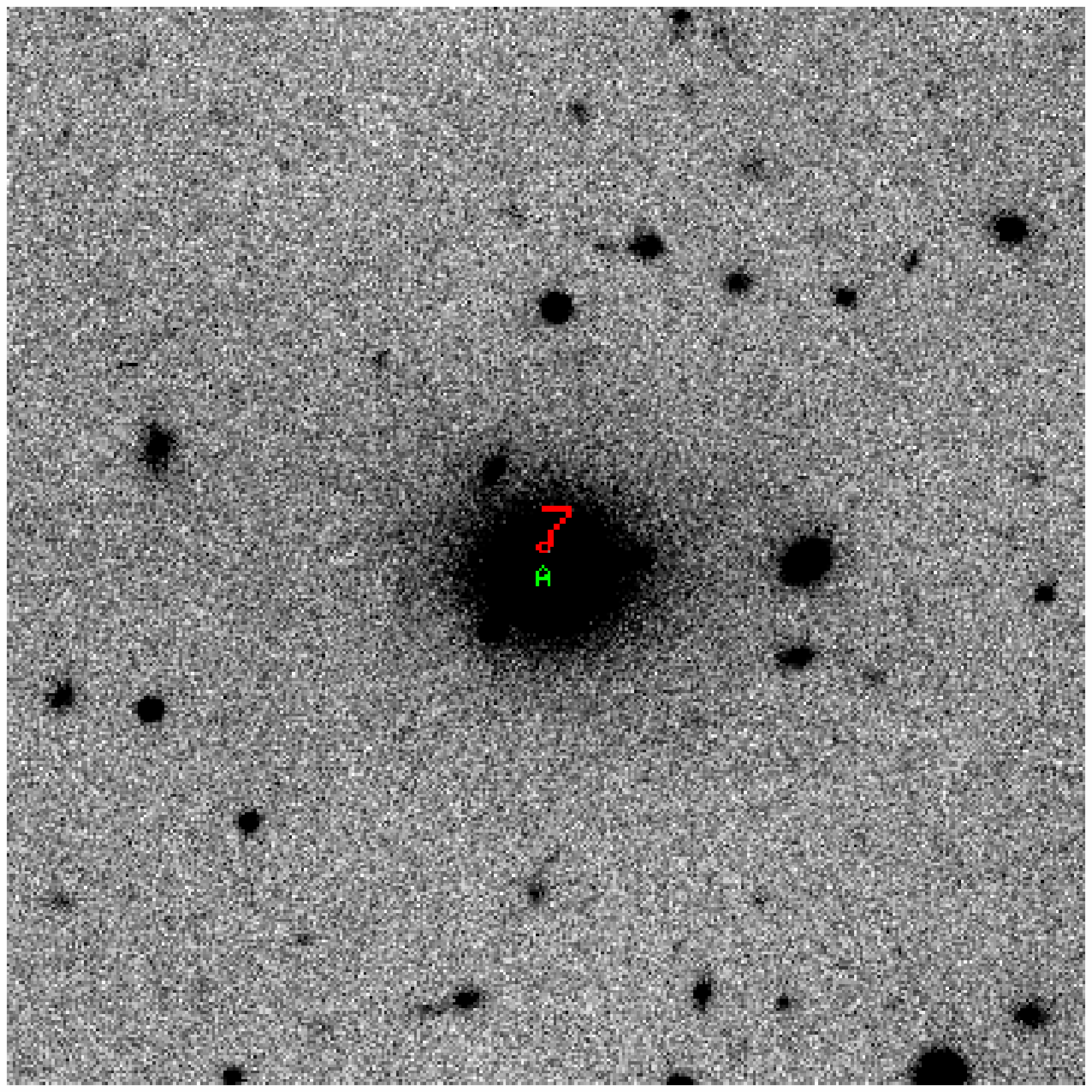}}
{\includegraphics[height=8cm,angle=270]{dull-spec.ps}}
\caption{An X-ray luminous optically-dull galaxy found in the XMS
survey.  The image (r', $2'\times 2'$) was obtained with the WFC, the
red circle being the 1$\sigma$ statistical error circle of the X-ray
source. The spectrum is from the AUTOFIB2/WYFFOS
spectrograph on the WHT.}
\label{dull}
\end{figure}

Our survey also finds a number of X-ray sources which we identify as
galaxies without emission lines. In most cases there is real evidence
that these sources are directly associated with an optically dull
galaxy. One of such cases is shown in figure \ref{dull}. This is a
galaxy at z=0.095 with a 0.5-4.5 keV luminosity of $1.3 \times
10^{42}\, {\rm erg}\, {\rm s}^{-1}$, i.e., some sort of AGN must be
present in that galaxy to give rise to such a high luminosity,
although there is no hint of optical emission lines. The nature of 
such sources are discussed in Severgnini et al (2002). We must
caution, however, that other cases where the putative identification
of the X-ray source is an optically dull galaxy, might actually
correspond to a group or cluster of galaxies.  Careful inspection of
both the optical images and the possible extent of the X-ray sources
is then necessary.

\section{The AGN distribution}

\begin{figure}[h]
\resizebox{\hsize}{!}
{\includegraphics[height=8cm, angle=270]{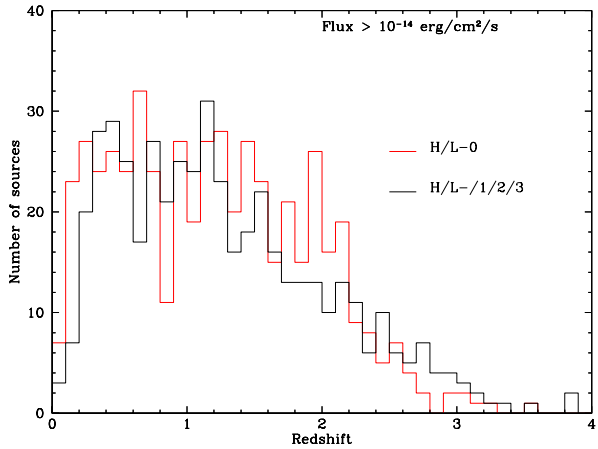}}
\caption{Expected redshift distribution for the AGNs in the Comastri
et al (1995) model for the X-ray background, for a 0.5-4.5 keV flux
limited sample at $10^{-14}\, {\rm erg}\, {\rm cm}^{-2}\, {\rm
s}^{-1}$. The red line shows the distribution of unabsorbed sources
and the black line the distribution of absorbed sources, including
both high /H) and low (L) luminosity AGNs.}
\label{types}
\end{figure}

Synthesis models of the X-ray background based on the AGN unified
scheme, make specific predictions for the redshift and absorption
column distribution of AGN. We show in fig. \ref{types} the expected
redshift distribution in the Comastri et al (1995) model, as drawn from a simulation.

Similar numbers of absorbed and unabsorbed objects should be found at
a flux level $\sim 10^{-14}\, {\rm erg}\, {\rm cm}^{-2}\, {\rm
s}^{-1}$. Also, the redshift distributions for both absorbed and
unabsorbed AGN are expected to be similar in this model.  Indeed both
of these predictions depend critically on the assumptions of the XRB
model and in fact they would be different in, e.g., the Gilli, Salvati
\& Hasinger (2001) model, in the sense that more absorbed AGN should
be found at high redshifts.

\begin{figure}[h]
\resizebox{\hsize}{!}
{\includegraphics{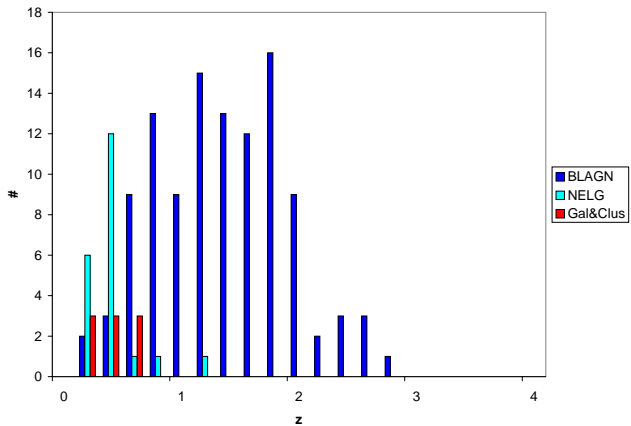}}
\caption{Redshift distribution for the various classes of
extragalactic objects found in the XMS survey.}
\label{distz}
\end{figure}

For comparison, figure \ref{distz} shows the redshift distribution for
the various classes of extragalactic objects of the sources identified
so far in this survey. If we add all of them (or even adding the
broad-line and narrow-line AGN), we find rough agreement between the
observed and predicted redshift distributions of the Comastri et al
(1995) model. The problem arises if the usual identification of X-ray
unabsorbed with type 1 and X-ray absorbed with type 2 X-ray to optical
link is assumed.  Clearly type 2 AGN are much more concentrated at low
redshift than type 1 AGN.  This appears to be a feature of even the
deepest X-ray surveys, and therefore it presents an important
challenge to the XRB synthesis models.  Possible ways to understand
this mismatch include

\begin{itemize}

\item Type 1 and type 2 AGN are truly independent populations with
unrelated evolution stories. Along these lines Franceschini, Braito \&
Fadda (2002) propose that type 1 AGN at high redshifts form at the
highest and rarest peaks of the density field in the Universe, while
type 2 AGN are more closely related to the assembly of large galaxies
at $z\sim 1$, and are therefore more related to star formation.

\item The XRB synthesis models make assumptions about the X-ray
absorption of AGNs, not about optical spectral types.  There is a
wealth of evidence for discordant cases in the literature, including
X-ray absorbed type 1 AGNs and X-ray unabsorbed type 2 AGNs. Mateos et
al (2003) discuss the possibility that type 1 AGN at high z are
absorbed in the X-ray domain.

\end{itemize}

\section{Broad Absorption Line QSOs}

Four Broad Absorption Line QSOs have been found so far in our X-ray
selected survey, one of them below the flux threshold (see table
\ref{BAL}).  Such objects have been so far extremely rare in X-ray
surveys (Green \& Mathur 1996), the likely explanation being that the BAL gas absorbs X-rays
in such a way that the X-ray flux decreases significantly.  Indeed, 4
out of 110 appears to be less than the expected fraction of BAL QSOs
in an optically selected sample (typically 10\%), but in a
representative number of cases we do not have enough wavelength
coverage to detect BALs.  

\begin{table}[h]
\caption{Parameters for the 4 X-ray selected BAL QSOs. S is the
0.5-4.5 keV flux and BI is the {\it
Balnicity Index} (in ${\rm km}\, {\rm s}^{-1}$), as defined in Weymann
et al (1991). The BI marked with $^{\star}$ is uncertain.}
\label{BAL}
\begin{tabular}{c c c}\hline
$z_{QSO}$ & $S ({\rm erg\, cm^{-2}\, s^{-1}})$ & BI \\ 
\hline
1.820 & $2.96\times 10^{-14}  $ & 3650 \\
1.919 & $2.52\times 10^{-14}  $ & 580  \\
1.881 & $2.46\times 10^{-14}  $ & 1620$^{\star}$ \\
2.000 & $1.61\times 10^{-14}  $ & 1460 \\
\hline
\end{tabular}
\end{table}

We must stress that all 4 BALs belong to the high ionisation class, as
the absorption detected is in the CIV line. The {\it Balnicity Index}
(Weymann et al 1991) measures the strength of the (CIV)
absorption. Compared to the Weymann et al (1991) sample, it appears
that the BI of these X-ray selected BAL QSOs is on the low to medium
side. This, together with the possible high ionisation of the BAL,
might explain why they are detected in X-rays.

\section{Outlook}

Pending completion of the identification of about 1/3 of the 239 X-ray
sources in the XMS, the picture emerging at a source density of $\sim
100\, {\rm deg}^{-2}$ is that Broad-Line AGN dominate the content of
the X-ray sky.  The redshift distribution of these objects is very
similar at that depth to that of deeper surveys (Lehmann et al 2001),
which probably means that for studies of the bulk of the
high-luminosity population of broad-line AGN there is no need to go
much deeper. Narrow-line AGN appear only at much lower redshifts in
this survey (which is also being found in the deepest surveys),
meaning that the combination of XRB synthesis models plus a one-to-one
correspondence of AGN X-ray absorption with optical spectral type,
does not work. This highlights the need for re-thinking current
popular interpretations of X-ray background models.

\acknowledgements
This paper reports on work carried out in the framework of the
XMM-Newton Survey Science Centre. We are grateful to our consortium
colleagues and especially to those involved in the AXIS
project. Partial financial support for this research was provided by
the Spanish Ministry of Science and Technology under projects
AYA2000-1690 (XM, FJC, MTC, SM) and PB98-0409-C02-01 (IPF).

\end{document}